\documentclass[aps,prl,showpacs,floatfix,twocolumn,amsmath,amssymb,preprintnumbers]{revtex4-1}
\usepackage{mathrsfs}
\usepackage[figuresright]{rotating}
\usepackage{amsmath}
\usepackage{amssymb}
\usepackage{graphicx}% Include figure files
\usepackage{color}
\usepackage{dcolumn}% Align table columns on decimal point
\usepackage{bm}% bold math
\usepackage[breaklinks=true,colorlinks=true,linkcolor=blue,urlcolor=blue,citecolor=blue]{hyperref}

\usepackage{soul}

\makeatletter

\newcommand{\Rmnum}[1]{\expandafter\@slowromancap\romannumeral #1@}
\makeatother

\begin{document}

%\title{Skyrmion Depinning at Ultra-Low Current Densities}
\title{Elasticity in the skyrmion phase unveils depinning at ultra-low current densities}

\author{Yongkang Luo$^{1}$}
\email[]{mpzslyk@gmail.com}
\author{Shizeng Lin$^{1}$}
\author{M. Leroux$^{1}$}
\author{N. Wakeham$^{1}$}
\author{D. M. Fobes$^{1}$}
\author{E. D. Bauer$^{1}$}
\author{J. B. Betts$^{1}$}
\author{J. D. Thompson$^{1}$}
\author{A. Migliori$^{1}$}
\author{M. Janoschek$^{1}$}
\author{Boris Maiorov$^{1}$}
\email[]{maiorov@lanl.gov}
\affiliation{$^1$ Los Alamos National Laboratory, Los Alamos, New Mexico 87545, USA.}

\date{\today}

\begin{abstract}
Controlled movement of nano-scale stable magnetic objects has been proposed as the foundation for a new generation of magnetic storage devices. Magnetic skyrmions, vortex-like spin textures stabilized by their topology are particularly promising candidates for this technology. Their nanometric size and ability to be displaced in response to an electrical current density several orders of magnitude lower than required to induce motion of magnetic domain walls suggest their potential for high-density memory devices that can be operated at low power. However, to achieve this, skyrmion movement needs to be controlled, where a key question concerns the coupling of skyrmions with the underlying atomic lattice and disorder (pinning). Here, we use Resonant Ultrasound Spectroscopy (RUS), a probe highly sensitive to changes in the elastic properties, to shed new light on skyrmion elasticity and depinning in the archetypal skyrmion material MnSi. In MnSi, skyrmions form a lattice that leads to pronounced changes in the elastic properties of the atomic lattice as a result of magneto-crystalline coupling. Without an applied current, the shear and compressional moduli of the underlying crystal lattice exhibit an abrupt change in the field-temperature range where skyrmions form. For current densities exceeding $j_c^*$ the changes of elastic properties vanish, signaling the decoupling of skyrmion and atomic lattices. Interestingly, $j_c^*$, which we identify as the onset of skyrmion depinning, is about 20 times smaller than $j_c$ previously measured via non-linear Hall effect. Our results suggest the presence of a previously-undetected intermediate dynamic regime possibly dominated by skyrmion-creep motion with important consequences for potential applications.
\end{abstract}

\maketitle

%\section{Significance}
%High packing densities and with ultra-low current motion position skyrmions as next-generation memory and logic devices, a primary motivation for this work. Skyrmions are a swirl of spins that form close packed lattices. Using Resonant Ultrasound Spectroscopy (RUS), we present the first study of the response of the complete elastic tensor to the formation of skyrmions in MnSi and find that changes in shear moduli are weaker than those for compression moduli upon entry into the skyrmion phase. These results do not agree with models but are similar to measurements on superconducting vortex lattices. We find that decoupling of skyrmions from the lattice occurs above $j_c^*\sim 20-60~$\,kA/m$^{2}$, which is 20 times lower than measured by other techniques. This is the first application of an ultrasonic method to detect skyrmion depinning, and could prove decisive for the practical application of skyrmion systems where the Hall effect changes are too small to be detected.

\section{\Rmnum{1}. Introduction}

Magnetic skyrmions are topologically-stabilized objects with a whirl-like spin-texture that emerges from a magnetic field- and temperature-tuned balance of magnetic interactions\cite{Muhlbauer-SKX}. Their response to electrical currents as well as their nanoscale size make them highly promising for applications in spintronics and memory devices\cite{Fert,Mueller-2track}. Because of their topological properties, they pin weakly to the underlying atomic lattice and defects. This results in critical current densities required to induce skyrmion motion that are ultra-low compared to currents typically necessary to produce current-driven magnetization dynamics in magnetic domains, making them, among others, prime candidates for so-called race-track memory devices. All proposed devices architectures require a detailed understanding of skyrmion pinning and in some cases on being able to engineer appropriate pinning landscapes to control skyrmion movement.

\begin{figure}[htbp]
\hspace*{-8pt}
\vspace*{-20pt}
\includegraphics[width=9.8cm]{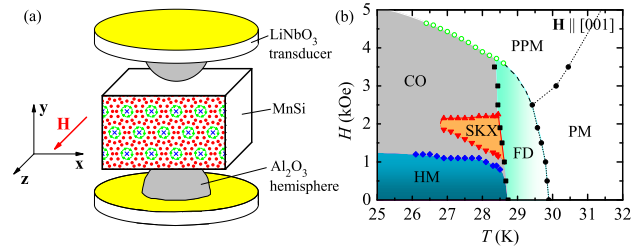}
%\vspace*{10pt}
\caption{\label{Fig1} (a) Schematic diagram of the RUS experimental setup. (b)  RUS magnetic phase diagram. The open circles are from $\chi'(H)$, defined as the mid-point of spin polarization. The abbreviations denoting the different phases are: HM = helimagnetic, CO = conical, SKX = skyrmion lattice, FD = fluctuation disordered, PM = paramagnetic, and PPM = polarized paramagnetic.}
\end{figure}

In bulk materials, a single skyrmion is a line-like structure oriented parallel to an external magnetic field (\textbf{H}) [see Fig.~\ref{Fig1}(a)] and in this sense is qualitatively analogous to vortices of magnetic flux that form in a superconductor subject to an applied magnetic field. Moreover, skyrmion lines also form a hexagonal skyrmion lattice (SKX) due to repulsive interactions and can be moved in response to an electromotive force sufficiently large to depin them from the disorder present in the sample. Because of these similarities, it is tempting to apply to skyrmions certain pinning models developed for superconducting vortices\cite{blatterrev}. The depinning current density for a skyrmion lattice in single crystals is low, indicating weak pinning, and results of previous studies have been successfully related to a weak collective pinning model\cite{Schulz-MnSiTHE,Jonietz-MnSiCurrent}. In this model, the critical current which sets the onset of movement is determined by a trade-off between the strength of the pinning potential and the elastic stiffness of the lattice\cite{blatterrev}. This critical current depends {\it inversely} on the lattice stiffness, meaning that a perfectly rigid lattice of linear objects like vortices or skyrmions cannot be pinned\cite{blatterrev}.

Theoretically, however, the interaction between pinning centers and skyrmions differs from  superconducting vortices, notably because skyrmion motion is dominated by a large Magnus force that tends to scatter skyrmions with a velocity perpendicular to the pinning force. This allows skyrmions to avoid passing through the pinning centers in contrast to vortices for which the Magnus force can be disregarded\cite{Lin-particle}. Thus, for example, skyrmions that depin because of thermal fluctuations cannot be easily recaptured by pinning centers, making pinning of skyrmions less effective compared to vortices\cite{Lin-particle}.

Microscopically, skyrmion movement has been investigated directly through Lorentz Transmission Electron Microscopy (LTEM)\cite{Yu-FeGe} and by the spin-transfer torque over the lattice seen by small-angle neutron scattering (SANS)\cite{Jonietz-MnSiCurrent}. Macroscopically, it has also been inferred from a reduction of the topological Hall effect (THE) at large current densities\cite{PinningSKX,dong2015}. However, apart from the associated macroscopic critical current density, $j_c$, no detailed experimental information on the depinning process is available, and it remains unknown how and which material defects affect skyrmions and to what extent their spin influences pinning\cite{Lin-particle}.

We exploit this magneto-crystalline coupling by using Resonant Ultrasound Spectroscopy (RUS) measurements to determine with unprecedented resolution the elastic response of the atomic lattice to the formation and depinning of the skyrmion lattice in the prototypical skyrmion lattice material MnSi. From the changes of the measured bulk elastic properties, we can infer the skyrmion lattice elastic properties and detect depinning of skyrmions. Quantitative values of SKX elasticity are useful in constructing an effective skyrmion pinning theory, and determining the relevant parameters needed for successful pinning models.  Previous ultrasound measurements on MnSi and related MnGe based on a pulse-echo technique have already demonstrated the sensitivity to the presence of the SKX as a consequence of its coupling to the crystalline lattice, revealing diverse behavior depending on the type of skyrmion lattice (MnSi, MnGe etc)\cite{Nii-MnSiElastic,KanazawaMnGe,Petrova-MnSiCij}. Skyrmion lattice stiffness was estimated to be 3 orders of magnitude smaller than that of the atomic lattice of MnSi, being as soft as the vortex lattice in superconductors\cite{Nii-MnSiElastic}. The relative small variation in elastic constants, permits the use of the changes in bulk stiffness as a measure of the skyrmion lattice stiffness, at least as a first order\cite{Nii-MnSiElastic}.

Compared to pulsed-echo measurements, RUS used here has significant advantages; RUS determines the complete elastic tensor in a single measurement with no transducer-sample bond\cite{Migliori-RUS,Migliori-PhysicaB1993}. By measuring all the elastic moduli $C_{ij}$ {\it simultaneously} with a fixed magnetic field orientation, accounting for anisotropic demagnetization factors is unnecessary and direct comparisons among $C_{ij}$ can be made. In contrast, pulse-echo studies require different samples or rotating the sample with respect to the magnetic field to obtain the complete elastic tensor\cite{Nii-MnSiElastic,Petrova-MnSiCij}.

Our RUS measurements reveal the complete elastic tensor (six independent irreducible moduli $C_{ij}$) of a single crystal of MnSi and the coupling of the SKX to the bulk elastic response of the atomic lattice. Upon entering the skyrmion phase, we find a much smaller response for shear than for compressional moduli, confirming a similarity with superconducting vortex lattices\cite{blatterrev}. Applying current, we identify the elastic response related to the decoupling of the SKX above a critical current density $j_c^*$. We determine the temperature dependence of $j_c^*$ and compare it to the thermodynamic stiffness variation associated with the skyrmion lattice measured with zero current applied, determining the limits of applicability of the weak collective pinning model. The value of $j_c^*$ is $\sim$20 times smaller than the critical current density $j_c$ measured by changes in THE and SANS\cite{PinningSKX,Jonietz-MnSiCurrent}. Our work identifies $j_c^*$ is the onset of skyrmion movement, and that the differences in magnitude and temperature dependence with respect to $j_c$ suggests the presence of an intermediate dynamic regime dominated by SKX creep motion that was previously undetected.

%\vspace*{-35pt}

\section{\Rmnum{2}. Experimental details}

The MnSi single crystal was grown by the Bridgman-Stockbarger method followed by a 1-week anneal at 900$^\circ$C in vacuum. The stoichiometry ratio of Mn:Si=0.88:1.12 was estimated by energy dispersive x-ray spectroscopy (EDS). The sample was polished into a parallelepiped along the [001] direction with dimensions 1.446$\times$0.485$\times$0.767 mm$^3$. The orientation of the crystal was verified by Laue X-ray diffraction within 1$^\circ$. Electrical resistivity and AC susceptibility measurements revealed a magnetic transition as expected at $T_c$=28.7 K, and a residual resistance ratio $RRR$[$\equiv$$\rho(300~\text{K})/\rho(2~\text{K})$]=87, indicating a high quality single crystal [See Fig.~S1 in \textbf{Supporting information (SI)}]. All the measurements in this work were performed on the same crystal. Further, the SKX in a different piece of this sample was directly observed using SANS\cite{Fobes-MnSiStrain}.

A schematic diagram of the RUS setup is shown in Fig.~\ref{Fig1}(a). The sample was mounted between two LiNbO$_3$ transducers. In order to stabilize the sample in a magnetic field and maintain RUS-required weak transducer contact, Al$_2$O$_3$ hemispheres (that also act as wear plates and electrical insulators) were bonded to each transducer. The external magnetic field $\textbf{H}$ was applied along [001] of the cubic crystal structure of MnSi. Frequency sweeps from 1250 to 5300 kHz were performed for each measurement. The resonance peaks were tracked and recorded as a function of temperature and magnetic field. Elastic moduli $C_{ij}$ were extracted from 24 resonance frequencies with an RMS error of 0.2\% using an inversion algorithm\cite{Migliori-RUS,Migliori-PhysicaB1993}. Although the absolute error is large (mainly because of uncertainties in the sample dimensions), the precision of the elastic moduli determination is at least 1$\times$10$^{-7}$.

To study the effect of current on moduli, we attached gold wires (13 $\mu$m) at opposite sides of the specimen allowing a DC current to be applied along [100], \textbf{I}$\perp$\textbf{H} with a cross-section of 0.485$\times$0.767 mm$^2$. To minimize Joule heating at the contacts, the Au wires were spot-welded to the sample and covered with silver paint to improve current homogeneity and reduce contact resistance. The resulting Ohmic electrical contacts were less than 0.5 $\Omega$. Whenever the current was changed, we waited for steady state before recording. A small temperature increase ($<$30 mK) was observed at the thermometer right after applying relatively larger currents. We compensated for this by adjusting the set-point of the temperature controller.

\section{\Rmnum{3}. Results and discussions}

\begin{figure*}[htbp]
\vspace*{-10pt}
%\hspace*{-18pt}
\includegraphics[width=17cm]{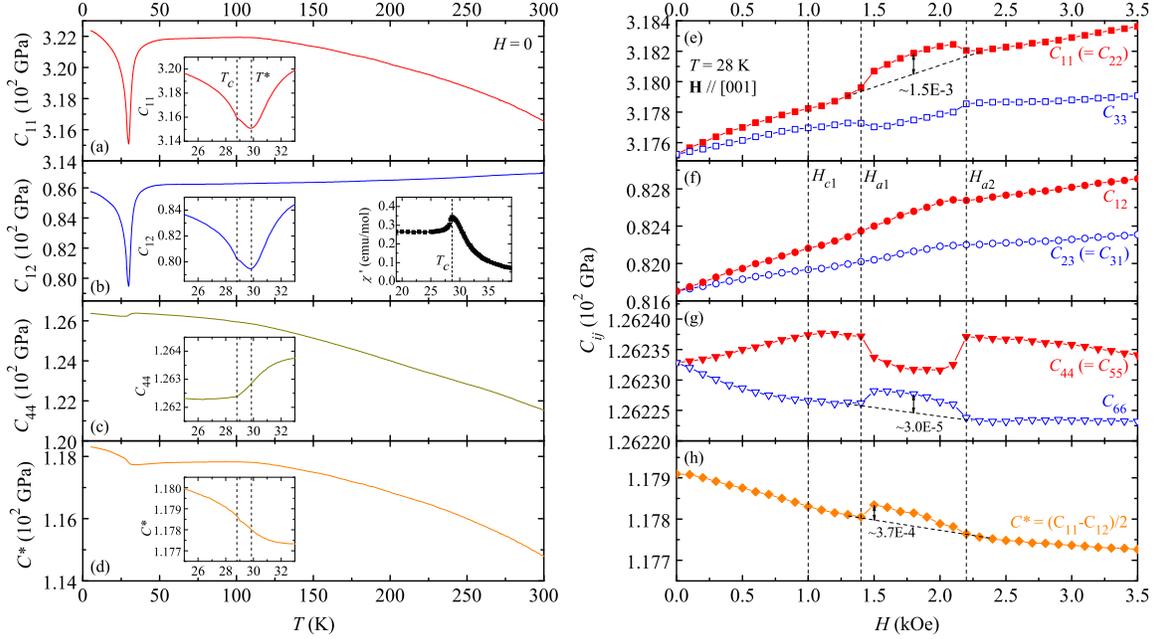}
\vspace*{-20pt}
\caption{\label{Fig2} Temperature dependence of elastic moduli of MnSi. (a-d) Zero-field temperature dependence of $C_{11}$, $C_{12}$ $C_{44}$ and $C^*$$\equiv$($C_{11}$$-$$C_{12}$)/2, respectively. The insets are near $T_c$=28.7 K. The right inset to panel (b) displays the real part of AC susceptibility $\chi'$ as a function of $T$. (e-h) Isothermal field dependent $C_{ij}$ at $T$=28 K.}
\end{figure*}

In this section, we describe the response of the elastic moduli to the formation of the SKX as a function of temperature ($T$) and magnetic field ($H$). In addition, we also explore the change of the elastic properties when electrical currents of density $j$ are applied to induce motion of the SKX. We use the extreme sensitivity of RUS to detect all relevant magnetic phase transitions, and the symmetry sensitivity of RUS to understand the observations. We reveal the envelope of skyrmion depinning, analyze the temperature dependence of critical depinning current $j_c^*$ in the framework of a weak collective pinning model, and provide insight into magneto-elastic coupling in this system.

Elastic moduli are thermodynamic susceptibilities that, when magneto-elastic coupling is important, are particularly favorable for observing phase boundaries. Details of the data analysis are described in \textbf{SI}. Figure \ref{Fig1}(b) illustrates the magnetic phase diagram obtained via the elastic moduli and supported by magnetic susceptibility measurements. Both give results in good agreement with the literature\cite{Muhlbauer-SKX} where the SKX phase appears in a narrow range of temperatures and field within the conically (CO) ordered magnetic phase. Figures \ref{Fig2}(a-c) show the temperature dependencies of $C_{ij}$ of MnSi for $H$=0 that we used to define the phase boundaries. For the cubic symmetry of MnSi and in the absence of a magnetic field, only three independent elastic moduli are required. Here we follow the tradition of using $C_{11}$, $C_{12}$ and $C_{44}$ as the independent elastic moduli (\textbf{SI}). In addition, the [110] shear modulus is defined as $C^{*}$$\equiv$$(C_{11}$$-$$C_{12})/2$.
%Because of no measurable corresponds to $C_{12}$, we also use the [110] shear modulus defined as $C^{*}$$\equiv$$(C_{11}-C_{12})/2$. As result of this, throughout the paper, we show $C_{12}$ for completeness but also show and discuss $C^*$.
At room temperature, we find $C_{11}$=316.54~GPa, $C_{12}$=86.96~GPa, and $C_{44}$=121.53~GPa in good agreement with the literature\cite{Petrova-MnSiCij}. Upon cooling, both $C_{11}$ and $C_{44}$ increase. As temperature is lowered below 35 K, $C_{11}$ and $C_{44}$ drop but $C^{*}$ rises. This is characteristic of softening in one symmetry direction and stiffening in another in the vicinity of the magnetic phase transition to the helimagnetic (HM) state, highlighting the importance of measuring the full elastic tensor. The change of $C_{44}$ and $C^{*}$ are much smaller than those of $C_{11}$ and $C_{12}$. $C_{11}$ exhibits a minimum at $T^*$=29.8 K before recovering and then displaying an inflection at $T_c$=28.7 K where the system undergoes a PM-HM (helimagnetic) transition. The derived $T_c$ agrees well with AC susceptibility measurements shown in the right inset to Fig.~\ref{Fig2}(b). In the low temperature limit, all $C_{ij}$ tend to saturate, as expected. It should be mentioned that $T^*$ is about 1 K above $T_c$, corresponding to a so-called Vollhardt invariance previously determined via specific heat and neutron scattering measurements\cite{Bauer-MnSi_FeCo, Janoschek-MnSi2013}. Notably, $T^*$ is a characteristic temperature that describes a crossover from mean-field ferromagnetic to strongly-interacting HM fluctuations. Thus, the window between $T^*$ and $T_c$ describes the fluctuation-disordered (FD) region on the phase diagram [Fig.~\ref{Fig1}(b)], where strong critical fluctuations of the helical order parameter reduce both the correlation length and the mean-field helical phase transition temperature, resulting in a Brazovskii-type first-order transition at $T_c$ \cite{Janoschek-MnSi2013,Brazovskii-1975}.

In the presence of a magnetic field $\textbf{H}$$\parallel$[001], symmetry is broken and the $\textbf{z}$-axis is no longer equivalent to the $\textbf{x}$- and $\textbf{y}$-axes. This lowers the symmetry of the elastic tensor from cubic to tetragonal (\textbf{SI}), requiring three additional independent elastic moduli, $C_{33}$, $C_{23}$(=$C_{31}$) and $C_{66}$. In Figs.~\ref{Fig2}(e-g), we plot $C_{ij}$ as a function of $H$ at $T$=28 K. Each subset of $C_{ij}$ splits into two branches under magnetic field. By extracting the six $C_{ij}$ from the same frequency scan, we are able to track all the moduli as a function of field and observe a discontinuous jump in some $C _{ij}$ between $H_{a1}$=1.4 kOe and $H_{a2}$=2.2 kOe. Based on the field dependence of $\chi'$ shown in Fig.~S2(b) (\textbf{SI}), we determine $H_{a1}$ and $H_{a2}$ as the lower and upper boundaries of the SKX phase, respectively. Below $H_{a1}$, there is a weak inflection in $C_{ij}(H)$ near $H_{c1}$=1.0 kOe, which we assign to the field-induced helimagnetic (HM)-conical (CO) phase  transition. A closer look at Figs.~\ref{Fig2}(e-g) reveals contrasting behavior for different $C_{ij}$. Although both compression ($C_{11}$, $C_{33}$) and shear moduli ($C_{44}$, $C_{66}$) display abrupt changes, $C_{12}$ and $C_{23}$ only exhibit a slight change in slope near $H_{a1}$ and $H_{a2}$. The amplitudes of the elastic moduli jumps are consistent with what was measured previously\cite{Petrova-MnSiCij,Nii-MnSiElastic}. A much smaller variation is found in the shear than in the compression moduli, being 3$\times$10$^{-5}$ and 1.5$\times$10$^{-3}$ GPa respectively [see Figs.~\ref{Fig2}(e,f)]. It is important to remember that we are measuring the elastic moduli of the entire crystal and although the changes are associated with the existence and properties of the SKX, the detected signal is characteristic of the SKX coupled to the overall sample elasticity. Looking at $C^*$ in Fig.~\ref{Fig2}(h), the jump is an order of magnitude bigger than that found for $C_{66}$, indicating that hexagonal symmetry of the SKX does not describe the system, as for this we would expect $C^*$=$C_{66}$ (\textbf{SI}). Thus, tetragonal symmetry resulting from applying a magnetic field to the cubic system is the appropriate symmetry to consider for this system.

Because we have computed the contribution of each elastic moduli to each resonance frequency, it is useful and more precise to plot the raw frequencies when only trends and discontinuities are sought\cite{Evans2017}. Table S1 in \textbf{SI} shows the resulting least-square-fitting for $C_{ij}$ measured at $T$=34 K and $H$=0. The dependence of each resonance frequency on each $C_{ij}$, i.e. $\partial F/\partial C_{ij}$, is presented (normalized) and shown in the last nine columns of the table. Because the third and fourth frequencies in the table depend predominantly on $C_{11}$, we can use these resonances, $F_{1654}$ and $F_{2419}$, as a simpler measure of $C_{11}$. For a frequency that depends only on one $C_{ij}$ we have $C_{ij}$$\propto$$F^2$, so for small changes of those $C_{ij}$ we have $\delta C_{ij}$$\sim$2$\delta F$.

\begin{figure}[htbp]
%\vspace*{-10pt}
\hspace*{-8pt}
\includegraphics[width=9cm]{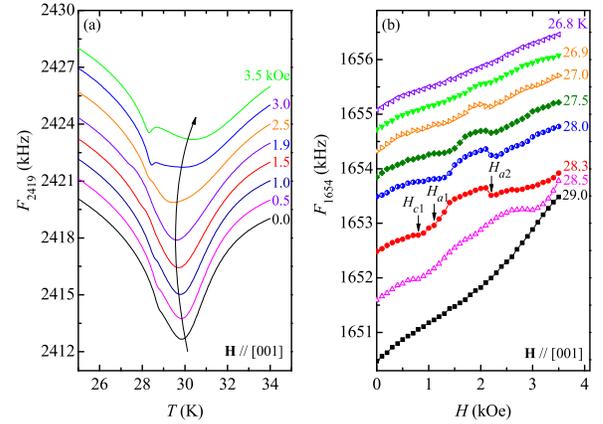}
\vspace*{-15pt}
\caption{\label{Fig3} Two representative RUS resonance-peak frequencies dominated by $C_{11}$. The curves are vertically shifted for clarity. (a) Temperature dependence of $F_{2419}$ at different fields. The arrow marks the evolution of the minimum of $F_{2419}$ with field. (b) Field dependence of $F_{1654}$ at various temperatures. Arrows mark characteristic fields corresponding to phase transitions.}
\end{figure}

\begin{figure*}[htbp]
\vspace*{-0pt}
\hspace*{-0pt}
\includegraphics[width=18cm]{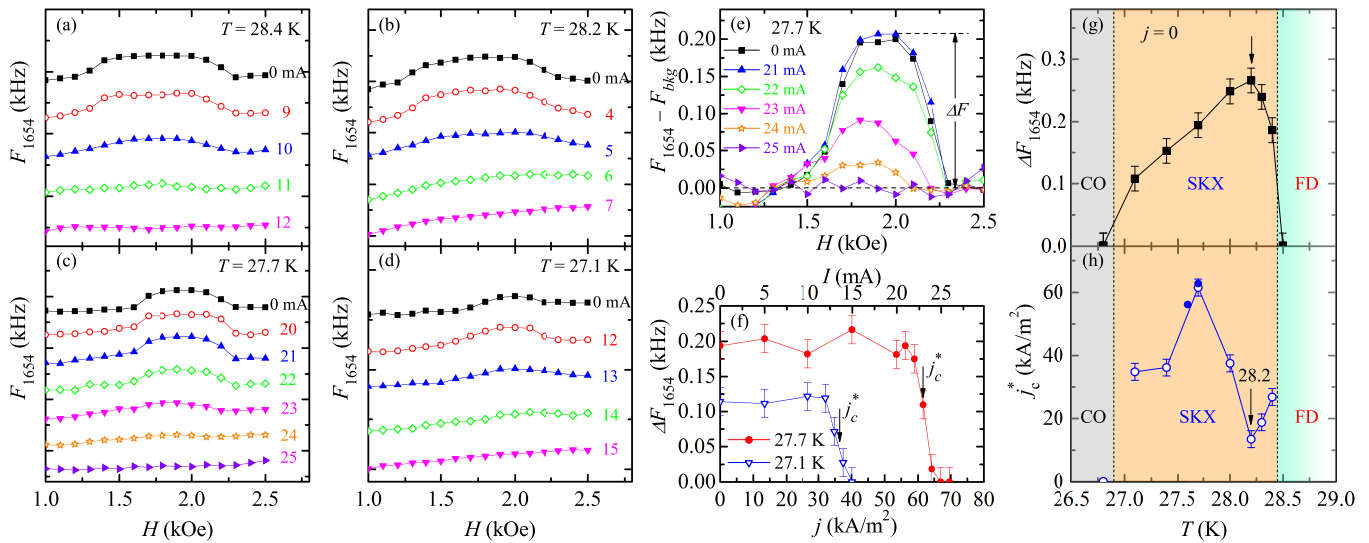}
\vspace*{-0pt}
\caption{\label{Fig4} Effect of applied current on $F_{1654}$. (a-d) Field dependence of $F_{1654}$ at various currents, measured at 28.4 K, 28.2 K, 27.7 K and 27.1 K, respectively. The curves are vertically offset. (e) $F_{1654}$$-$$F_{bkg}$ as a function of $H$ at 27.7 K, where $F_{bkg}$ is the smooth background of $F_{1654}$. $\Delta F$ is defined as the maximum of $F_{1654}$$-$$F_{bkg}$. (f) $\Delta F_{1654}$ vs. $j$ for 27.7 K and 27.1 K. The arrows mark the critical current density values $j_c^*$=62 and 35 kA/m$^2$, respectively. (g) $\Delta F_{1654}$ as a function of $T$ when in the absence of current. (h) temperature dependent $j_c^*$. The full symbols are measured with and ultrasonic excitation two time larger than that used to measured the open symbols.}
\end{figure*}

Using only resonance frequencies, Fig. \ref{Fig3}(a) displays the temperature dependence of the relevant elastic moduli at various magnetic fields. For $H$=0, the behavior of $F_{2419}(T)$ resembles that of $C_{11}(T)$, confirming the dominance of $C_{11}$. With increasing magnetic field, $T_c$ is gradually suppressed, and the signature of the phase transition becomes more pronounced for $H$$>$2.5 kOe. The temperature where minimum of $F_{2419}(T)$ occurs initially decreases with increasing $H$ but then broadens and shifts to higher $T$ for $H$$>$2.5 kOe where spins become polarized by the external field. Figure~\ref{Fig3}(b) shows $F_{1654}$ as a function of $H$ measured at selected temperatures. The discontinuous jump in $F_{1654}(H)$ can be identified between 26.9 K and 28.5 K similarly as seen in $C_{11}$ in Fig.~\ref{Fig2}(b). A positive jump in $F_{1654}(H)$ signifies stiffening in $C_{11}$ when the system enters the SKX phase. The magnitude of the jump in $F_{1654}(H)$, denoted by $\Delta F$, is plotted as a function of temperature in Fig.~\ref{Fig4}(g) with a maximum near the SKX-FD boundary and decreasing as $T$ decreases. The value of $\Delta F$ can be regarded as a measure of convolution between the SKX lattice stiffness and the strength of the coupling between SKX and the atomic lattice. Indeed, the maximum in $\Delta F$ near the upper boundary of the SKX ($T$$\sim$28.2 K) phase can be explained by proximity to the fluctuation disordered regime; the abundance of critical magnetic fluctuations makes the skyrmion lattice softer, which would allow for better pinning to material disorder (see below).

Electrical current also influences elastic properties of the atomic lattice in the SKX phase. Figures~\ref{Fig4}(a-d) display the field dependence of $F_{1654}$ with various applied electrical currents (\textbf{I}$\perp$\textbf{H}) at four selected temperatures 28.4, 28.2, 27.7 and 27.1 K inside the SKX phase, respectively. Taking $T$=27.7 K as an example (Fig.~\ref{Fig4}c) , we subtract a smooth background $F_{bkg}(H)$ from the raw $F_{1654}(H)$, the results of which are shown in [Fig.~\ref{Fig4}(e)]. $\Delta F$ is then defined as the maximum of $F_{1654}$$-$$F_{bkg}$. $\Delta F$ remains essentially unchanged for current values up to 21 mA, drops abruptly between 22 and 24 mA, and becomes nearly unresolvable at 25 mA. The threshold current density $j_c^*$ is defined as the midpoint of the drop in $\Delta F$, shown in Fig.~\ref{Fig4}(f) with $j_c^*$=62(3) kA/m$^2$ for 27.7 K. The error bar is set by the step size in current. In Fig.~\ref{Fig4}(f) we also show $\Delta F(j)$ for 27.1 K, and the difference in $j_c^*$s is far larger than the measurement uncertainty. The $F_{1654}(H)$ curves for $j$$>$$j_c^*$ do not display the characteristic stiffening signature corresponding to the skyrmion phase, as if the SKX is completely decoupled from the lattice.

We emphasize that the changes observed in the resonance frequencies are not caused by current-induced Joule heating as can be illustrated by several observations. (i) The magnitude of $\Delta F$ should increase with $j$ [cf Fig.~\ref{Fig4}(f)] for temperatures near the lower boundary of the SKX phase (e.g. 27.1 K). (ii) $j_c^*$ should increase monotonically with decreasing temperature. (iii) For significant Joule heating, the first point in Fig.~\ref{Fig4}(g), $T$=26.8 K below the skyrmion phase boundary, should display a non-zero $\Delta F$, in contrast to what it is measured. (iv) The phase boundaries in $H$ depend strongly on temperature as revealed in the phase diagram in Fig.~\ref{Fig1}(b). The width of the SKX phase with respect to field showing non-zero $\Delta F$ remains unaffected for increasing current density at all measured temperatures, as expected for constant temperature [Figs.~\ref{Fig4}(a-e)]. All these observations demonstrate that the coupling between the SKX and the crystals is lost at $j_c^*$.

The drastic changes in the elastic properties above $j_c^*$ recall the measurements of the critical current for the onset of the skyrmion movement\cite{Schulz-MnSiTHE,Jonietz-MnSiCurrent,Yu-FeGe,dong2015}. However, the critical current $j_c$$\sim$1 MA/m$^2$ derived from the decrease in the Hall signature\cite{Schulz-MnSiTHE}, which has been associated with the onset of movement of skyrmions caused by spin-torque\cite{Jonietz-MnSiCurrent}, is about 20 times larger than our RUS results display.

There are several scenarios that can explain the difference in values between $j_c$ and $j_c^*$. First, the presence of ultrasonic waves could facilitate depinning by shaking skyrmions off pinning potentials resulting in a smaller current for motion threshold. Such kind of $j_c$ decrease was observed in other pinned systems like superconducting vortices\cite{Valenzuela2001} and recently in MnSi when applying an alternating magnetic field\cite{privatecomm}. However, this effect is unlikely here because the same $j_c^*$ is observed with ultrasonic excitations increased by a factor of 2 [full symbols in Fig.~\ref{Fig4}(g)].

Another possibility is that the disorder is lower or of a different kind compared to the crystals used for previous studies, which would yield a lower $j_c$ in our sample. Because the $RRR$ of our sample is comparable to that of previous studies, disorder density seems unlikely to be the cause of the smaller $j_c$ observed here. However, the amount of disorder is not the only variable affecting skyrmion pinning (and order parameter); notably, the type, spin and valence of the defects need to be taken into account. Numerical simulations have shown that different types of defects may act as pinning or depinning centers depending on their nature\cite{Lin-particle, Choi2016}. In this context, we find that our single crystals exhibit a slight Mn deficiency as verified by EDS (\textbf{SI}). This is also reflected in measurable differences between the MnSi crystal studied here and those used for previous Hall effect studies. For example the $T_c$ here is about 0.5 K higher, and the Hall effect measured on several of our single crystals (including a slice taken from the same batch of the RUS sample) also shows differences, the details of which, will be part of a future publication. This points to a situation  where the type of disorder present in our sample allows for more efficient SKX depinning.

We also consider the different sensitivities for detecting the onset of skyrmion motion using diverse techniques. For Hall effect measurements, $j_c$ can only be identified after the motion of skyrmions produces a measurable change in the small THE. The situation in SANS is similar, as the resolution in reciprocal space will be limited if the skyrmions moves too slowly. In contrast, RUS directly measures the magneto-crystalline coupling, and thus detects skyrmion movement immediately when the SKX decouples (and hence a lower $j_c$).  Therefore, RUS is similar to LTEM experiments which detect ON-OFF movement. This scenario is supported by the contrasting current dependence between RUS and Hall effect responses used to determine $j_c$. The abruptly change in $\Delta F$ as $j$ reaches $j_c^*$ is distinct from the gradual decrease of the topological Hall resistivity for $j$$>$$j_c$ \cite{Schulz-MnSiTHE}.

A similar disparity occurs in detecting the critical current for superconducting vortex movement, in which $j_c$ values determined from magnetization measurements  are lower than those extracted using transport. The former are sensitive to creep and the latter to flux-flow changes\cite{blatterrev,borisnatmat,campbellevettsnew}. It is likely that the situation here is similar, where the differing temperature dependences observed for $j_c$ and $j_c^*$ suggest that RUS and the Hall effect sense different dynamic regimes. As skyrmion movement is dominated by the Magnus force, this allows thermal fluctuations to depin the SKX with ease. Notably, the escape rate due to thermal fluctuations is proportional to $e^{(-\Delta U/k_BT)}$, where $\Delta U$ is the height of the pinning potential\cite{Lin-particle}. However, the applied current may not be sufficient to induce direct movements of the SKX, but driven creep motion would dominate this intermediate dynamic regime. We note that neither Hall effect nor SANS measurements are sensitive to such incoherent SKX motion.

Finally, the temperature dependence of $j_c^*$ [Fig.~\ref{Fig4}(h)] provides additional important insight into skyrmion pinning. According to the weak collective pinning theory, $j_c^*$ is inversely proportional to the SKX stiffness. The temperature dependence of $\Delta F$($j$=0) [see Fig.~\ref{Fig4}(g)] displays two trends. Starting at high temperature from the SKX-FD, $\Delta F$ increases as $T$ decreases becoming stiffer down to $T$=28.2 K marked with an vertical arrow in Fig.~\ref{Fig4}(g). In agreement with weak collective pinning theory, we observe that $j_c^*$ decreases and shows a minimum at $T$=28.2 K [vertical arrow in Fig.~\ref{Fig4}(h)]. The behavior for $T$$>$28.2 K, is consistent with strong thermal fluctuations near the upper boundary of the SKX phase that soften the SKX lattice and allows skyrmion lines to accommodate better to disorder, improving pinning with the consequent higher $j_c^*$. The increase in $j_c^*$ near the SKX-FD phase boundary is called peak effect and was also observed in MnSi via the THE\cite{Schulz-MnSiTHE} and is ubiquitous to superconducting vortices in a weak pinning landscape\cite{blatterrev,Valenzuela2002}. As $T$ continues to decrease, $j_c^*$ and $\Delta F$($j$=0) keep displaying an inverse relation consistent with a weak collective pinning down to $T$=27.7 K where $j_c^*$ shows a maximum but $\Delta F$ only a weak inflection point. This temperature marks a change in pinning regime, indicating weak collective pinning is no longer valid below $T$=27.7 K. At lower temperatures $j_c^*$ decreases and then remains constant. No signature of a peak effect was observed near the lower-temperature phase boundary because the low temperature CO-SKX phase transition is not fluctuation dominated as is the SKX-FD. The non-monotonic $j_c^*(T)$ indicates that there are more than one pinning regimes.

\section{\Rmnum{4}. Conclusions}

In summary, we have determined the complete elastic tensor within and near the skyrmion phase of MnSi. Obtaining the full elastic tensor enables direct comparison of the relation between compression and shear moduli to understand the coupling of the SKX to the underlying crystalline lattice. The thermodynamic susceptibilities measured by RUS inform measurements obtained under dynamic conditions and suggest the presence of a previously undetected intermediate regime of skyrmion motion above critical current densities that are much smaller than determined by Hall effect and SANS measurements. This new intermediate regime is likely dominated by thermally induced creep motion of skyrmions that was predicted theoretically, but not yet observed. This suggests that skyrmion motion is possible at lower current densities than previously thought, and is an important step towards skyrmion-based applications. Our results also demonstrate that ultrasound detection is a new way to study skyrmion motion, in particular for materials where changes in the THE are small. Finally, combining high-density electrical current and ultrasonic techniques will be a useful technique to study other electric field and current sensitive systems such as multiferroic and superconducting materials.

\section{Acknowledgments}

We thank F.~F. Balakirev for technical support, and F. Ronning and M. Garst for insightful conversations. Work at Los Alamos National Laboratory (LANL) was performed under the auspices of the U.S. Department of Energy. Research by YL, SL, DMF, ML, ND, EDB, MJ and BM was supported by LANL Directed Research and Development program. Work by JB and AM was part of the Materials Science of Actinides, an Energy Frontier Research Center funded by the U.S. DOE, Office of Science, BES under Award DE-SC0001089.

%\bibliographystyle{apsrev}
%\bibliography{bibliobm}

\newpage

\setcounter{table}{0}
\setcounter{figure}{0}
\setcounter{equation}{0}
\renewcommand{\thefigure}{S\arabic{figure}}
\renewcommand{\thetable}{S\arabic{table}}
\renewcommand{\theequation}{S\arabic{equation}}
\onecolumngrid

\newpage

\begin{center}
{\bf \Large
{\it Supporting Information:}\\
%Resonant ultrasound spectroscopy: a new access to the elasticity of skyrmion lattice
%Elasticity of MnSi with skyrmion lattice in static and dynamic conditions
%Switching skyrmion elastic coupling via electrical currents: a Resonant ultrasound spectroscopy study}
%Depinning currents and elasticity of skyrmion lattice measured via Resonant Ultrasound Resonance
%Elasticity of the skyrmion lattice unveils physics of depinning at ultra-low currents
%Skyrmion Depinning at Ultra-Low Current Densities}
Elasticity in the skyrmion phase unveils depinning at ultra-low current densities}

\end{center}

\begin{center}
Yongkang Luo$^{1*}$\email{ykluo@lanl.gov}, Shizeng Lin$^{1}$, M. Leroux$^{1}$, N. Wakeham$^1$, D. M. Fobes$^{1}$,  E. D. Bauer$^{1}$, J. B. Betts$^{1}$, A. Migliori$^{1}$, J. D. Thompson$^1$, M. Janoschek$^{1}$, and Boris Maiorov$^{1\dag}$\email{maiorov@lanl.gov}\\
$^1${\it Los Alamos National Laboratory, Los Alamos, New Mexico 87545, USA.}\\

\date{\today}
\end{center}

In this \textbf{Supporting Information (SI)}, we provide additional energy dispersive x-ray spectroscopy (EDS), resistivity and AC susceptibility data. We also give a brief introduction of elastic moduli and details of data analysis used for RUS measurements that support the results, discussions and our conclusions given in the main text.

\section{\textbf{SI \Rmnum{1}: S\lowercase{ample characterization}}}

\begin{figure*}[htbp]
\hspace*{-10pt}
\vspace*{-20pt}
\includegraphics[width=17cm]{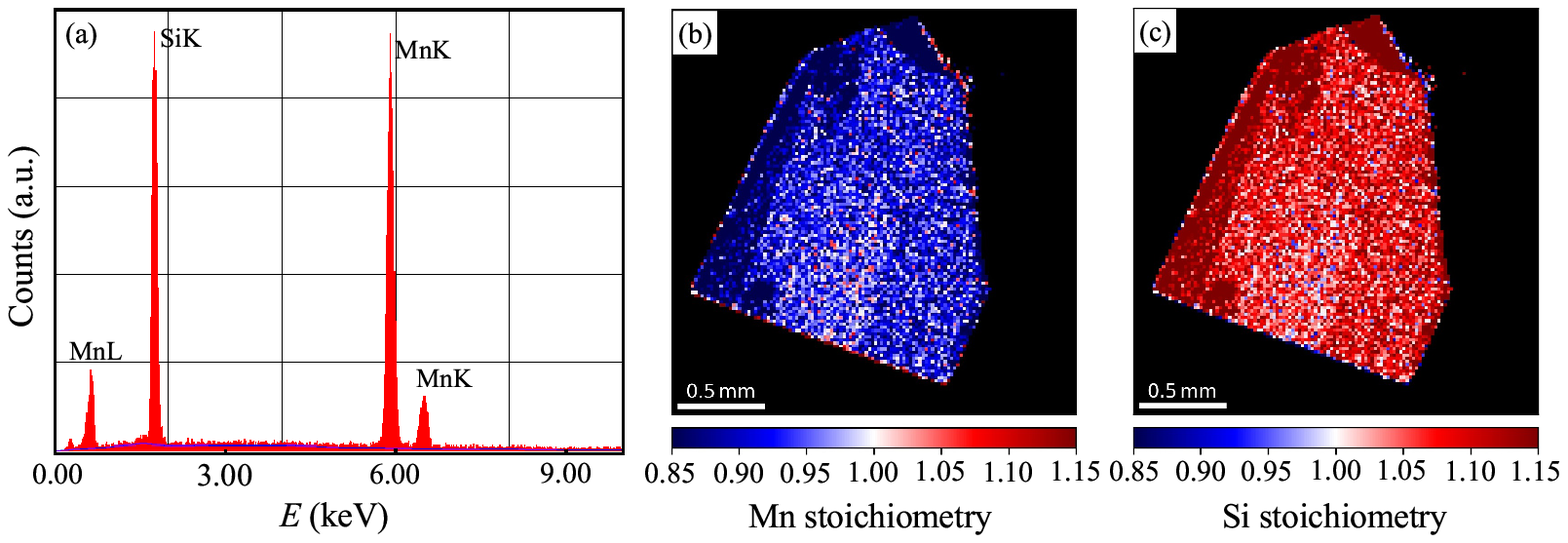}
\vspace*{-0pt}
\caption{\label{FigS1} (a) A representative EDS spectrum of MnSi. (b) and (c) show the maps of stoichiometric concentration of Mn and Si, respectively. The average mole ratio Mn:Si=0.88:1.12. }
\end{figure*}

The stoichiometric concentrations of Mn and Si of the sample were determined by EDS measurements. Figure \ref{FigS1}(a) shows a representative EDS spectrum. The distribution maps of Mn and Si are given in Fig.~\ref{FigS1}(b) and (c), respectively. Apparently, our single crystal exhibits a slight Mn deficiency. The average mole ratio Mn:Si=0.88:1.12. This Mn deficiency plays a crucial role in reducing the helical transition $T_c$ and SKX pinning, as discussed in main text.

\begin{figure*}[htbp]
\vspace*{-20pt}
\includegraphics[width=17cm]{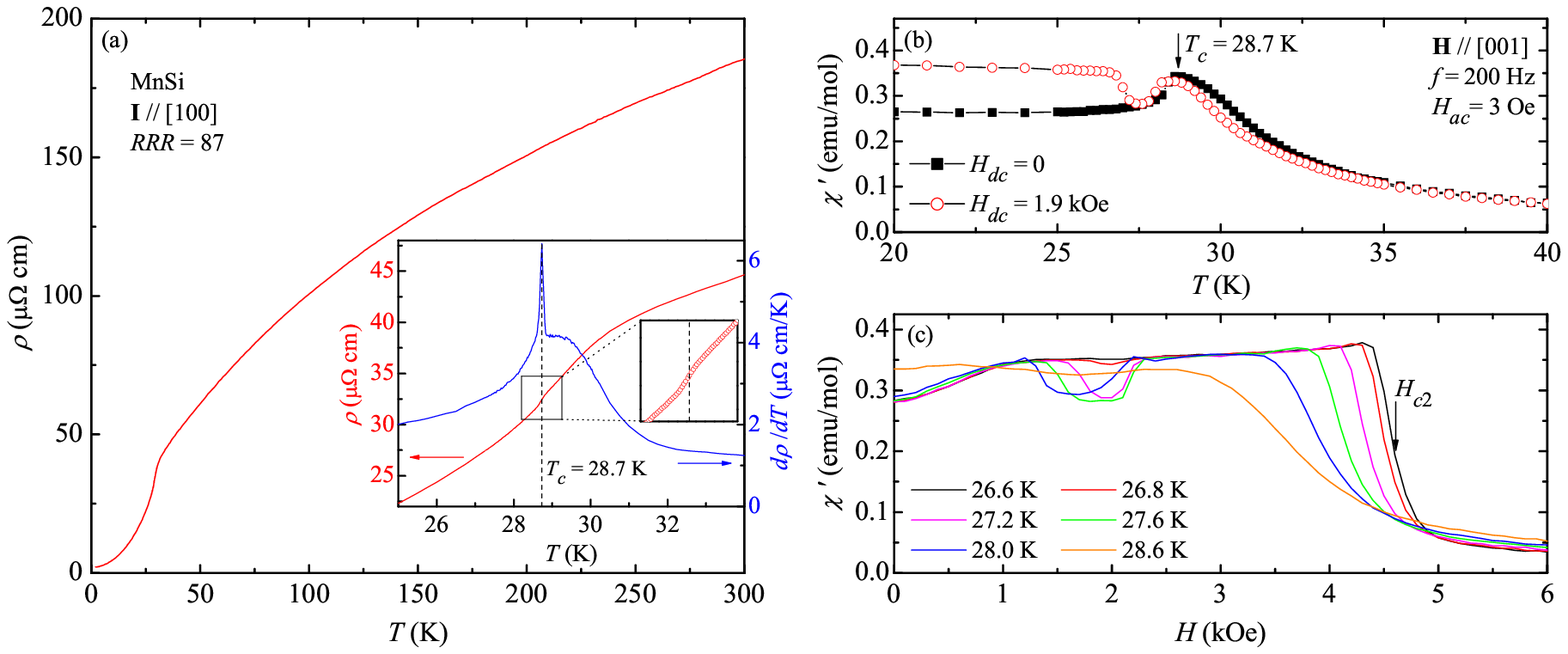}
\vspace*{-20pt}
\caption{\label{FigS2} (a) Temperature dependence of resistivity. The inset shows an enlarged view of $\rho(T)$ as well as $d\rho/dT$ near $T_c$=28.7 K. (b) Temperature dependent AC susceptibility $\chi'$, measured under DC magnetic field $H$=0 (black) and 1.9 kOe (red). (c) $\chi'(H)$ measured at various temperatures. The SKX phase appears in the window 26.8 K$<$$T$$<$28.6 K and between 1.5 and 2.3 kOe. }
\end{figure*}

Figure \ref{FigS2}(a) shows the temperature dependence of resistivity $\rho$ measured with current applied along [100]. The resistivity decreases sub-linearly with decreasing $T$. Below 32 K, $\rho(T)$ starts to decrease rapidly. An inflection point is observed at 28.7 K [inset to Fig.~\ref{FigS2}(a)], used for determining  $T_c$=28.7 K resistively. In the inset to Fig.~\ref{FigS2}(a), we also display $d\rho/dT$ as a function of $T$. A sharp peak in $d\rho/dT$ is found at $T_c$ on top of a broad maximum which extends to $\sim$32 K. Similar behavior was also seen in other thermodynamic properties\cite{Bauer-MnSiC,Stishov-MnSi2007}, and is a signature of pre-formed short-range HM order above $T_c$ and a fluctuation-induced first-order transition at $T_c$ \cite{Janoschek-MnSi2013}.

In Fig.~\ref{FigS2}(b) we present the results of AC susceptibility measured with an alternating field $H_{ac}$=3 Oe and frequency $f$=200 Hz. When $H_{dc}$=0, $\chi'(T)$ displays a pronounced peak at $T_c$, consistent with resistivity and RUS measurements. For $H$=1.9 kOe, $\chi'(T)$ shows a pronounced valley between 26.8 and 28.6 K characteristic of SKX phase. This SKX phase is also seen in the isothermal $\chi'(H)$ curves shown in Fig.~\ref{FigS2}(c). These observations are akin to earlier literature\cite{Bauer-MnSiPhase,Nii-MnSiElastic}, except for the slight different field range for SKX which is probably due to a different demagnetization factor.

\section{\textbf{SI \Rmnum{2}: $C\lowercase{_{ij}}$ \lowercase{and problem symmetry}}}

For a 3D crystal, we write Hooke's law as\cite{Kittel-SolidState,Migliori-RUS}
\begin{equation}
\boldsymbol{\sigma}=\mathbf{C}\cdot\boldsymbol{\varepsilon},
\label{EqS1}
\end{equation}
where $\boldsymbol{\sigma}$ is the stress tensor (6$\times$1), $\boldsymbol{\varepsilon}$ is the strain tensor (6$\times$1), and $\mathbf{C}$ is the elastic moduli matrix (6$\times$6). Expanding Eq.~(\ref{EqS1}),
\begin{equation}
\left[
 \begin{array}{c}
    \sigma_{1} \\
    \sigma_{2} \\
    \sigma_{3} \\
    \sigma_{4} \\
    \sigma_{5} \\
    \sigma_{6} \\
  \end{array}
\right]=
\left[
  \begin{array}{cccccc}
    C_{11} & C_{12} & C_{13} & C_{14} & C_{15} & C_{16}  \\
    C_{21} & C_{22} & C_{23} & C_{24} & C_{25} & C_{26}  \\
    C_{31} & C_{32} & C_{33} & C_{34} & C_{35} & C_{36}  \\
    C_{41} & C_{42} & C_{43} & C_{44} & C_{45} & C_{46}  \\
    C_{51} & C_{52} & C_{53} & C_{54} & C_{55} & C_{56}  \\
    C_{61} & C_{62} & C_{63} & C_{64} & C_{65} & C_{66}  \\
  \end{array}
\right]\left[
  \begin{array}{c}
    \varepsilon_{1} \\
    \varepsilon_{2} \\
    \varepsilon_{3} \\
    \varepsilon_{4} \\
    \varepsilon_{5} \\
    \varepsilon_{6} \\
  \end{array}
\right].
\label{EqS2}
\end{equation}
Note that in this notation the subscripts 1 to 6 are actually two-element subscripts, and they correspond to the Cartesian coordinates by the following decoder table:
\begin{equation}
1\leftrightarrow xx; ~~~2\leftrightarrow yy; ~~~3\leftrightarrow zz; ~~~4\leftrightarrow yz,zy; ~~~5\leftrightarrow zx,xz; ~~~6\leftrightarrow xy,yx.
\label{EqS3}
\end{equation}
Because $C_{ij}$=$C_{ji}$, there are 21 different elastic moduli for an arbitrary system. The higher symmetry of the system, the lower the number of independent elastic moduli in $\mathbf{C}$. For cubic symmetry, $\mathbf{C}$ becomes
\begin{equation}
\mathbf{C}=\left[
 \begin{array}{cccccc}
    C_{11} & C_{12} & C_{12} & 0      & 0      & 0       \\
    C_{12} & C_{11} & C_{12} & 0      & 0      & 0       \\
    C_{12} & C_{12} & C_{11} & 0      & 0      & 0       \\
    0      & 0      & 0      & C_{44} & 0      & 0       \\
    0      & 0      & 0      & 0      & C_{44} & 0       \\
    0      & 0      & 0      & 0      & 0      & C_{44}  \\
  \end{array}
\right],
\label{EqS4}
\end{equation}
i.e., there are only three independent elastic moduli, $C_{11}$, $C_{12}$ and $C_{44}$. This is the case for MnSi in its paramagnetic state for $H$=0. The ground state of MnSi becomes a helimagnetic order below $T_c$, which in principle breaks the cubic symmetry. Treating MnSi as tetragonal for $H=0$ and $T<T_c$ is beyond the scope of this work and will be part of future efforts.

If we apply an external magnetic field $\textbf{H}$$\parallel$[001], the principal $\textbf{z}$-axis is no longer equivalent to $\textbf{x}$- and $\textbf{y}$-axes, the system can be regarded as tetragonal symmetry, and therefore, we have three extra elastic moduli, $C_{33}$, $C_{23}$(=$C_{31}$) and $C_{66}$. This situation is applicable to MnSi in the polarized paramagnetic state when a magnetic field is applied.

The skyrmion lines typically form a hexagonal lattice. An additional constraint $C^*$$\equiv$$(C_{11}-C_{12})/2$=$C_{66}$ is imposed to a hexagonal symmetry, requiring now only five independent elastic moduli. This is not what we observe experimentally in Fig. 2, as $C^*$$\neq$$C_{66}$, indicating that the SKX is not setting the symmetry of the system that remains tetragonal. Furthermore, as the SKX is coupled to the crystalline lattice, this enables the observation of changes measured in the elastic moduli as shown in Fig. 2. These changes are small compared with elastic moduli from the crystalline lattice, allowing to treat the system as one with tetragonal symmetry. However, how these two lattices with different symmetries couple to each other and how to obtain a more accurate determination of the elastic moduli of SKX from the measured bulk moduli require further investigations.

The situation becomes more complicated when a DC current is applied along [100]. The current is expected to break the tetragonal symmetry into an orthorhombic one. In principle, there should be nine independent elastic moduli, $C_{11}$, $C_{22}$, $C_{33}$, $C_{12}$, $C_{23}$, $C_{31}$, $C_{44}$, $C_{55}$ and $C_{66}$. This level of difficulty is not granted as the experimental results show that the applied current does not change the resonance frequencies or their field or temperature dependence, indicating that the symmetry of the system is not broken by low applied currents.

\section{\textbf{SI \Rmnum{3}: D\lowercase{ata collection and analysis}}}

\begin{figure*}[htbp]
\includegraphics[width=15cm]{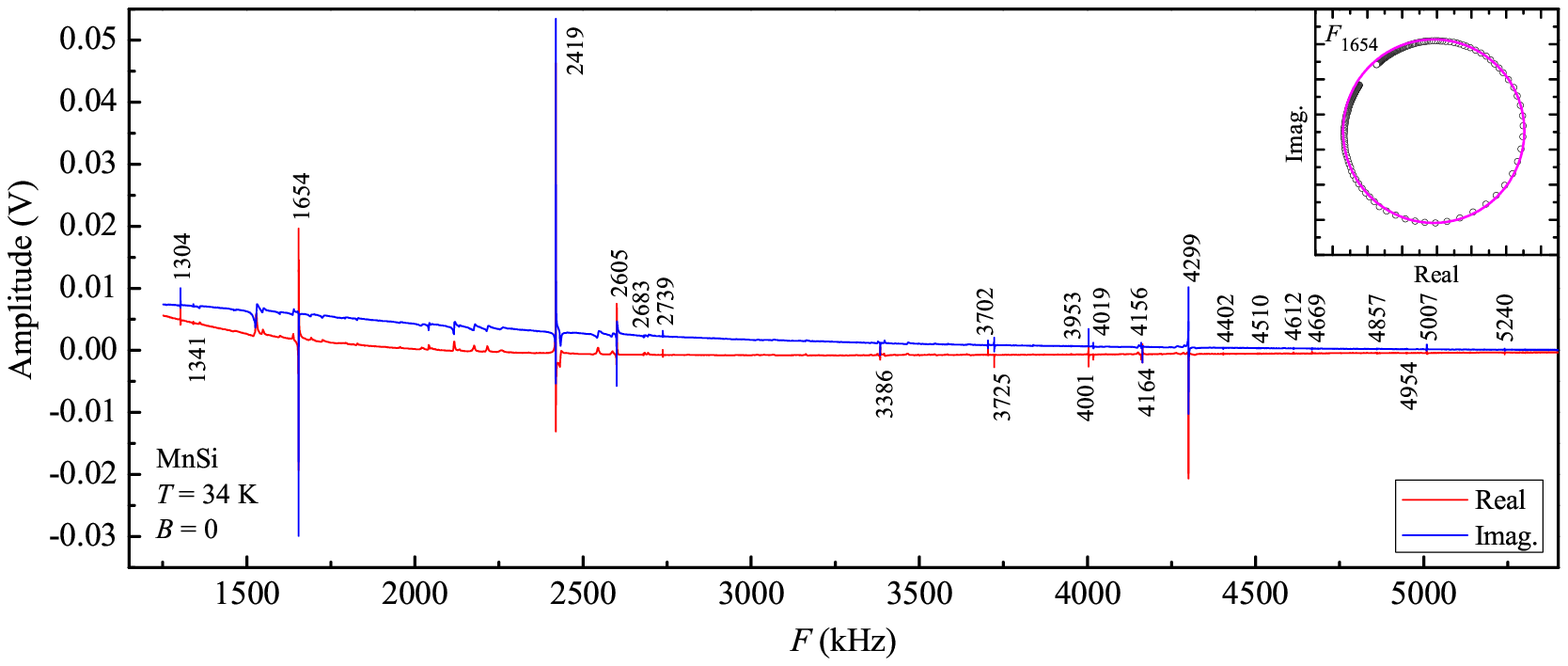}
\vspace*{-0pt}
\caption{\label{FigS3} Full RUS spectrum of MnSi collected at $T$=34 K in the absence of a magnetic field. The inset shows an Imaginary vs. Real plot of the resonance $F_{1654}$. The open symbols are experimental data, and the solid line defines a circle. }
\end{figure*}

In our RUS measurement, we swept frequency from 1250 to 5300 kHz. When frequency is approaching a normal mode of the sample, a resonance is detected. We show a representative RUS spectrum of MnSi in Fig.~\ref{FigS3} taken at $T$=34 K in the absence of magnetic field. For each resonance peak, the real part (in-phase, red) and imaginary part (out-of-phase, blue) of the signal have a Lorentzian shape $V({\it\omega})$=$z_0$$+$$A e^{i\phi}$/$({\it\omega}$$-$${\it\omega}_0$$+$$i{\it\Gamma}$$/$$2)$, where ${\it\Gamma}$ is the resonance width and $z_0$ is the background in the vicinity of the resonance. Plotting the real versus imaginary components of the resonance should form a circle in the complex plane (inset to Fig.~\ref{FigS3}). Both the width (${\it\Gamma}$) and resonance frequency ($\omega_0$) were obtained following an established approach\cite{Shekhter-Pseudo}.

For a sample with known elastic moduli, dimensions, and density, all the resonance peaks can be theoretically calculated by solving a 3D elastic wave function\cite{Migliori-RUS,Migliori-PhysicaB1993,Leisure-RUS}. In our case, the elastic moduli are derived by a least-square-fitting, in which $C_{ij}$  are set as free fitting parameters. The iteration continues until $\chi^2$$\equiv$$\sum_{n}[(F_{n}^{cal}$$-$$F_n^{exp})/F_n^{exp}]^2$ minimizes, where $F_n^{exp}$ and $F_n^{cal}$ are the $n$th experimental and calculated frequencies, respectively. Table \ref{TabS1} shows an example of this fitting for a measurement at $T$=34 K and $H$=0 for cubic symmetry. The degree to which each resonance frequency depends on different $C_{ij}$, i.e. $\partial F/\partial C_{ij}$, is then calculated (normalized) and shown in the last nine columns of the table.

\begin{table*}
\caption{\label{TabS1} The least-square-fitting of RUS peaks at $T$=34 K and $H$=0. The sample dimensions are 1.446$\times$0.485$\times$0.767 mm$^3$. The fitting derives $C_{11}$=$C_{22}$=$C_{33}$=3.204192$\times$10$^2$ GPa, $C_{12}$=$C_{23}$=$C_{13}$=0.849543
$\times$10$^2$ GPa, and $C_{44}$=$C_{55}$=$C_{66}$=1.263798$\times$10$^2$ GPa. $\partial F/\partial C_{ij}$ (normalized) characterizes the dependencies of the fitting of a resonance peak on each $C_{ij}$.}
\begin{ruledtabular}
\begin{center}
\def\temptablewidth{1.6\columnwidth}
\begin{tabular}{ccccccccccccc}
$n$ & $F^{exp}$ (kHz) & $F^{cal}$ (kHz) & $Err$ (\%) &    &    &    &  \multicolumn{3}{l}{~~~~~~~~~$\partial F/\partial C_{ij}$}   &    &    &         \\
    &                 &                 &            & $C_{11}$ & $C_{22}$ & $C_{33}$ & $C_{23}$ & $C_{31}$ & $C_{12}$ & $C_{44}$ & $C_{55}$ & $C_{66}$ \\
\hline
  1 & 1304.090 & 1295.876 & -0.63 &  0.95 &  0.03 &  0.04 &  0.02 & -0.10 & -0.09 &  0.00 &  0.00 &  0.14 \\
  2 & 1341.150 & 1338.273 & -0.21 &  0.00 &  0.00 &  0.00 &  0.00 &  0.00 &  0.00 &  0.02 &  0.69 &  0.28 \\
  3 & 1654.028 & 1653.957 &  0.00 &  0.93 &  0.01 &  0.03 &  0.02 & -0.08 & -0.10 &  0.00 &  0.19 &  0.00 \\
  4 & 2418.830 & 2417.748 & -0.05 &  1.11 &  0.05 &  0.07 &  0.03 & -0.14 & -0.11 &  0.00 &  0.00 &  0.00 \\
  5 & 2605.364 & 2606.384 &  0.04 &  0.10 &  0.01 &  0.01 &  0.00 & -0.01 & -0.01 &  0.10 &  0.56 &  0.24 \\
  6 & 2682.732 & 2688.634 &  0.22 &  0.69 &  0.02 &  0.05 &  0.01 & -0.09 & -0.05 &  0.00 &  0.00 &  0.37 \\
  7 & 2739.155 & 2745.038 &  0.21 &  0.31 &  0.02 &  0.01 &  0.00 & -0.02 & -0.04 &  0.00 &  0.71 &  0.00 \\
  8 & 3385.540 & 3380.757 & -0.14 &  0.03 &  0.04 &  0.84 & -0.08 & -0.05 &  0.01 &  0.20 &  0.00 &  0.02 \\
  9 & 3702.542 & 3699.256 & -0.09 &  0.02 &  0.03 &  0.70 & -0.07 & -0.03 &  0.00 &  0.16 &  0.06 &  0.12 \\
 10 & 3724.552 & 3725.367 &  0.02 &  0.26 &  0.01 &  0.03 &  0.00 & -0.04 & -0.02 &  0.23 &  0.30 &  0.24 \\
 11 & 3953.041 & 3986.015 &  0.83 &  0.30 &  0.02 &  0.04 &  0.00 & -0.02 & -0.04 &  0.00 &  0.70 &  0.00 \\
 12 & 4001.499 & 3995.519 & -0.15 &  0.20 &  0.07 &  0.84 & -0.11 & -0.10 &  0.00 &  0.00 &  0.09 &  0.00 \\
 13 & 4018.697 & 4018.478 & -0.01 &  0.34 &  0.02 &  0.04 &  0.01 & -0.03 & -0.04 &  0.00 &  0.68 &  0.00 \\
 14 & 4155.843 & 4140.983 & -0.36 &  0.18 &  0.08 &  0.78 & -0.11 & -0.08 &  0.00 &  0.00 &  0.15 &  0.00 \\
 15 & 4163.504 & 4181.133 &  0.42 &  0.49 &  0.02 &  0.08 &  0.00 & -0.05 & -0.04 &  0.02 &  0.02 &  0.46 \\
 16 & 4299.043 & 4298.411 & -0.01 &  0.58 &  0.00 &  0.74 & -0.02 & -0.32 &  0.01 &  0.00 &  0.00 &  0.00 \\
 17 & 4402.400 & 4379.128 & -0.53 &  0.25 &  0.01 &  0.06 &  0.00 & -0.05 & -0.02 &  0.40 &  0.05 &  0.30 \\
 18 & 4510.335 & 4513.721 &  0.08 &  0.03 &  0.03 &  0.45 & -0.04 & -0.01 &  0.00 &  0.09 &  0.14 &  0.32 \\
 19 & 4612.002 & 4609.303 & -0.06 &  0.02 &  0.15 &  1.04 & -0.19 & -0.02 &  0.00 &  0.00 &  0.00 &  0.00 \\
 20 & 4669.281 & 4648.204 & -0.45 &  0.03 &  0.01 &  0.07 &  0.00 & -0.01 &  0.00 &  0.52 &  0.08 &  0.31 \\
 21 & 4856.554 & 4854.090 & -0.05 &  0.00 &  0.01 &  0.13 & -0.01 &  0.00 &  0.00 &  0.86 &  0.00 &  0.00 \\
 22 & 4953.862 & 4960.544 &  0.13 &  0.12 &  0.01 &  0.02 &  0.00 & -0.02 &  0.00 &  0.00 &  0.01 &  0.88 \\
 23 & 5007.325 & 5020.570 &  0.26 &  0.74 &  0.25 &  0.19 & -0.09 &  0.11 & -0.20 &  0.00 &  0.00 &  0.00 \\
 24 & 5240.128 & 5251.871 &  0.22 &  0.49 &  0.07 &  0.58 & -0.05 & -0.17 & -0.03 &  0.00 &  0.11 &  0.00 \\
\end{tabular}
\end{center}
\end{ruledtabular}
\end{table*}

\end{document}